\documentclass{ptephy}

\usepackage{boites}
\newcommand{\br}{{\bf r}}
\usepackage{array}
\usepackage{natbib}

\begin{document}

\title{Tensor force and shape evolution of Si isotopes in Skyrme-Hartree-Fock model}

\author{\name{\fname{A.} \surname{Li}}{1,2,3,\ast}, \name{\fname{X. R.} \surname{Zhou}}{4,\ast}, and \name{\fname{H.}
\surname{Sagawa}}{1,5,\ast}}

\address{\affil{1}{RIKEN, Nishina Center, Wako 351-0198, Japan}
\affil{2}{Department of Astronomy and Institute of Theoretical
Physics and Astrophysics, Xiamen University, Xiamen 361005,
China}\affil{3}{State Key Laboratory of Theoretical Physics,
Institute of Theoretical Physics, Chinese Academy of Sciences,
Beijing, 100190}\affil{4}{Department of Physics and Institute of
Theoretical Physics and Astrophysics, Xiamen University, Xiamen
361005, China} \affil{5}{Center for Mathematics and Physics,
University of Aizu, Aizu-Wakamatsu, Fukushima 965-8560, Japan}
\email{liang@xmu.edu.cn (AL); xrzhou@xmu.edu.cn (XZ);
sagawa@u-aizu.ac.jp (HS)}}

%\date{\today}

\begin{abstract}

Much interest has been devoted to the shape evolution of
neutron-rich Si isotopes recently both experimentally and
theoretically. We provide our study of $^{28-42}$Si by
Skyrme-Hartree-Fock model with BCS approximation for the pairing
channel. We use an empirical pairing parameter deduced from the
experimental binding energy data for each nuclei and each Skyrme
parametrization. The recent highlight of the deformation change from
a prolate shape of $^{38}$Si to an oblate shape of $^{42}$Si is
confirmed in the present model. We also emphasize the role of tensor
force on deformations of neutron-rich $^{30,32}$Si, and discuss its
underlying physics.

\end{abstract}

%\bigskip
%\pacs{
% 21.80.+a    % Hypernuclei
% 21.10.Dr    % Binding energies and masses
% 21.30.Fe    % Forces in hadronic systems and effective interactions
%     }
%\vskip1cm

\maketitle

%________________________________________________________________
\section{Introduction}

The disappearance of $N = 28$ shell closures has been a huge debate
in recent years, especially after the conflicting experimental
evidence on the structure of $^{42}$Si coming from MSU~\cite{Fri06}
and GANIL~\cite{Bas07}. The small two-proton removal cross section
in the first experiment was interpreted as a doubly closed-shell
structure with a large $Z = 14$ subshell gap at $N = 28$, indicating
a nearly spherical shape of this nucleus. Contrary to this result,
the observation of a very low-lying $2^+$ state at GANIL supports
the vanishing of the $N = 28$ shell closure and a largely-deformed
shape for $^{42}$Si. The later claim actually has been theoretically
pointed out in several studies with shell-model and mean-field
approaches~\cite{Wer96,Per00,Lal99,Guz02,Now09,Li11,Ots08,Uts12}.
These studies predict the spherical neutron shell closure is broken
already in the mean-field
level~\cite{Wer96,Per00,Lal99,Guz02,Now09,Li11}, and the shape
coexistence is present in some nuclei along the $N = 28$ isotone,
due to a soft energy surface~\cite{Wer96,Per00}. Some articles have
also considered the possible influence of the $\gamma$-softness
found in the neutron-rich Si region~\cite{Wer96,Per00,Uts12}.

In Ref.~\cite{Wer96}, comparisons were made among the
Skyrme-Hartree-Fock (SHF) approach, the relativistic mean-field
(RMF) model, and the finite-range droplet model (FRDM) for the
quadruple deformation of Si isotopes. These models gave consistent
ground-state shape predictions for $^{28}$Si ($N = Z = 14$),
$^{34}$Si ($N = 20, Z = 14$), and $^{42}$Si ($N = 28, Z = 14$):
oblate for $^{28,42}$Si and spherical for $^{34}$Si. The non-magic
nature of $^{42}$Si was noted then as an interesting phenomenon. The
reason for this is attributed recently to the tensor effect on the
change of shell structure in neutron-rich
nuclei~\cite{Ots08,Uts12,Uts09,Ots01,Ots05}. This tensor interaction
acts directly on the $1d$ spin-orbit splitting, and gives a smaller
$2s_{1/2}$-$1d_{5/2}$ proton gap when more neutrons occupy
$1f_{7/2}$ orbit. This feature is later analyzed in details by
Tarpanov et al.~\cite{Tar08} within various mean-field models
assuming the spherical symmetry. They have demonstrated that the
tensor component in the effective interaction governs the reduction
of the $1d$ proton spin-orbit splitting when going from $^{34}$Si to
$^{42}$Si. Also the magnitude of the change is consistent with the
empirical observations. It is necessary to extend the study of
tensor correlations not only for the spin-orbit splitting but also
for the evolution of deformed nuclei as will be done in the present
work. In addition, the inclusion of tensor terms in the calculations
have achieved considerable success in explaining several nuclear
structure problems not only in the ground
states~\cite{Col07,Bri07,Les07} but also in the excited
states~\cite{Bai09,Cao09,Bai10}. In the present study we focus on
the tensor effect on the shape evolution of Si isotopes, for
example: is the tensor-force-driven deformation present in other
neutron-rich Si isotopes, especially $^{30}$Si with a possible $N =
16$ subshell, since some models (for example, the FRDM~\cite{Mol95})
predicted a spherical for this nucleus~\cite{Wer96} while the large
$B(E2)$ value suggested its deformed nature~\cite{Ibb98}.

For this purpose, we use the deformed Skyrme-Hartree-Fock model
(DSHF)~\cite{Vau73} with BCS approximation for the nucleon pairing.
As well-known, the standard BCS treatment of pairing was criticized
to have the unphysical particle-gas problem~\cite{Naz94} when
applied to neutron-rich isotopes, namely, there is an unrealistic
pairing of highly excited states. To cure this defect to some
extent, a smooth energy-dependent cut-off weight was
introduced~\cite{Kri90,Ben00} in the evaluation of the local pair
density, to confine the region of the influence of the pairing
potential to the vicinity of the Fermi surface. Some work also tried
a variant calculation with weak pairing~\cite{Wer96,Li13}. In the
present study, the above-mentioned cut-off weight is incorporated in
the BCS approximation, and for each nucleus the pairing parameter is
obtained with full respect of the empirical data of pairing gaps,
extracted by using the experimental binding energies of
Ref.~\cite{Aud03} and the three-point mass difference
formula~\cite{Sat98}:
$\Delta^{(3)}(A)=\frac{(-)^A}{2}[B(A)+B(A-2)-2B(A-1)]$ with $B(A)$
the binding energy of nucleus with even mass number $A$. We employ
several recent Skyrme parametrizations with various tensor couplings
to explore the role of tensor force on the shape of neutron-rich Si
isotopes, and theoretical results will be confronted directly with
recent experiments.

The paper is organized as follows. In Sec. II, we outline the
necessary formalism. The numerical results and discussions are given
in Sec. III. Finally, Sec. IV contains the main conclusions of this
work.

%-------------------------------------------------------------------------------
\section{Formalism}

\begin{figure}
\centering
\includegraphics[width=13.5cm]{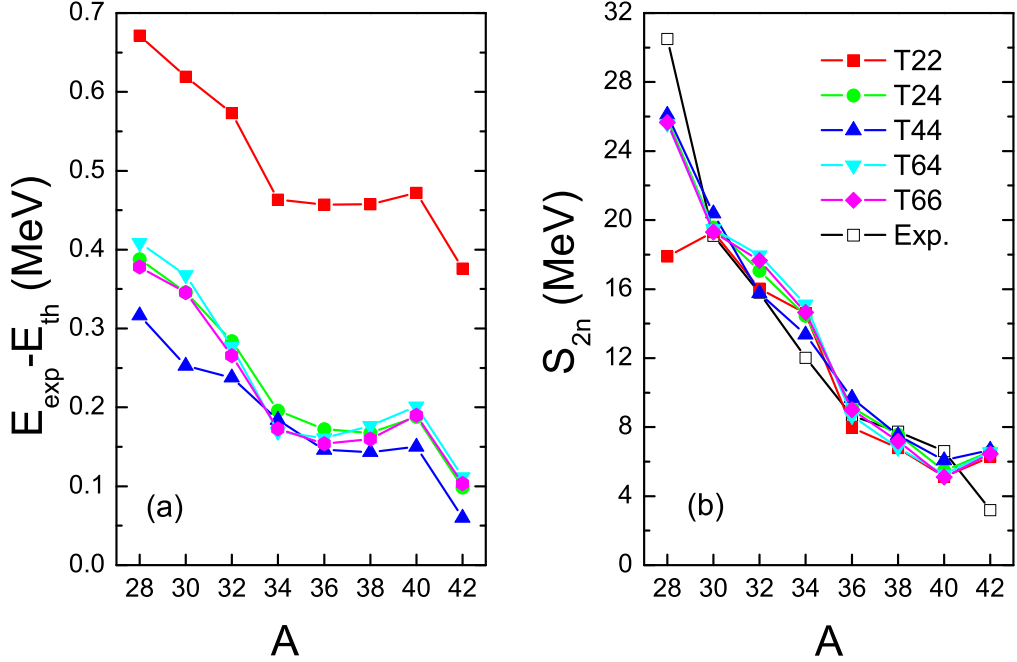}
\caption{(Color online) Left panel: the deviation of the theoretical
binding energies from the experimental values~\cite{Aud03} for
$^{28-42}$Si calculated with effective forces T22, T24, T44, T64,
and T66; Right panel: the two-neutron separation energies S$_{2n}$
in $^{28-42}$Si calculated with effective forces T22, T24, T44, T64,
and T66, in comparison with those of the experimental
data~\cite{Aud03}.}\label{fig1}
\end{figure}

We use the DSHF method~\cite{Vau73} solved in the coordinate space
with axially symmetric shape~\cite{Blum}. The pairing correlations
are taken into account by the BCS approximation with a
density-dependent pairing as done in Ref.~\cite{Sag04,Zho07,Li13}.
Further details can be found in the above references and are not
repeated here.

The tensor force is included in the same manner with in
Refs.~\cite{Li13,Bro06,Col07,Bri07,Les07}, with $T$ and $U$ denoting
the coupling constants representing the strength of the triplet-even
and triplet-odd tensor interaction, respectively. It contributes to
the Skyrme energy functional $H^{\text{Sk}}$ in a combined way with
the exchange term of central part as
\begin{eqnarray}
H_T^{\text{Sk}} &=& \frac{1}{2}\alpha(J_n^2+J_p^2) + \beta\vec{J}_n\cdot\vec{J}_p; \\
\alpha     &=& \alpha_C+\alpha_T;  \quad \beta = \beta_C+\beta_T  \label{eq:t} \\
\alpha_C   &=& \frac{1}{8}(t_1-t_2) - \frac{1}{8}(t_1x_1+t_2x_2); \nonumber \\
\beta_C    &=& -\frac{1}{8}(t_1x_1+t_2x_2) \label{eq:t1} \\
\alpha_T   &=& \frac{5}{12}U;  \quad \beta_T = \frac{5}{24}(T+U).
\label{eq:t2}
\end{eqnarray}
where subscripts $T$ and $C$ indicate the tensor and the central
contributions, respectively, and $J_q~(q=n,p)$ is the spin-orbit
density. $\alpha$ represents the strength of like-particle coupling
between neutron-neutron or proton-proton, and $\beta$ is that of the
neutron-proton coupling. With the contributions of tensor
interactions, the spin-orbit potential is given by,
\begin{eqnarray}
W_q=\frac{W_0}{2r}\left(2\frac{d\rho_q}{dr}+\frac{d\rho_{q'}}{dr}\right)+\left(\alpha\frac{
J_q}{r}+\beta\frac{J_{q'}}{r}\right). \label{Wq}\end{eqnarray} where
the first term comes from the Skyrme spin-orbit interaction whereas
the second one includes both the central exchange and tensor
contributions. The interactions between like (unlike) particles are
denoted $q (q')$ in Eq. \eqref{Wq}. Our selected parametrizations
are from the recently-built 36 effective interactions with
systematically adjusting of the tensor coupling
strengths~\cite{Les07}, which include: the Skyrme force T22 serving
as a reference with vanishing $J^2$ terms, T24 with a substantial
like-particle coupling constant $\alpha$ and a vanishing
proton-neutron coupling constant $\beta$; T44 with a mixture of
like-particle and proton-neutron tensor terms, T62 with a large
proton-neutron coupling constant $\beta$ and a vanishing
like-particle coupling constant $\alpha$; and T66 with large and
equal proton-neutron and like-particle tensor-term coupling
constants. The coupling strengths of various parameter sets used in
this study are collected in Tab.~1. To check the capability of the
above-selected effective forces with tensor terms, we display in
Fig.~\ref{fig1} the deviations of the calculated energies (left
panel) from the experimental data, and the predicted two-neutron
separation energies (right panel) in comparison with experiments for
$^{28-42}$Si. Experimental values are based on Ref.~\cite{Aud03}.
All the effective interactions overestimate the binding energy (less
than 0.7 MeV) along the isotopic chain. However, the interactions
with tensor terms (T24, T44, T64, and T66) give better descriptions
than the interaction T22 with the absence of tensor terms. This is
also true for the case of two-neutron separation energies S$_{2n}$.
All of T24, T44, T64, and T66 with the tensor terms are reasonably
well to describe the energy data S$_{2n}$. One should notice that
the two-neutron separation energies are not close to zero in Fig. 1,
more than 3 MeV even in $^{42}$Si. Thus we can conclude that all the
nuclei discussed in the present study are not loosely-bound nuclei,
which may justify the usage of BCS treatment for the pairing
channel.

\begin{figure}
\centering
\includegraphics[width=13.5cm]{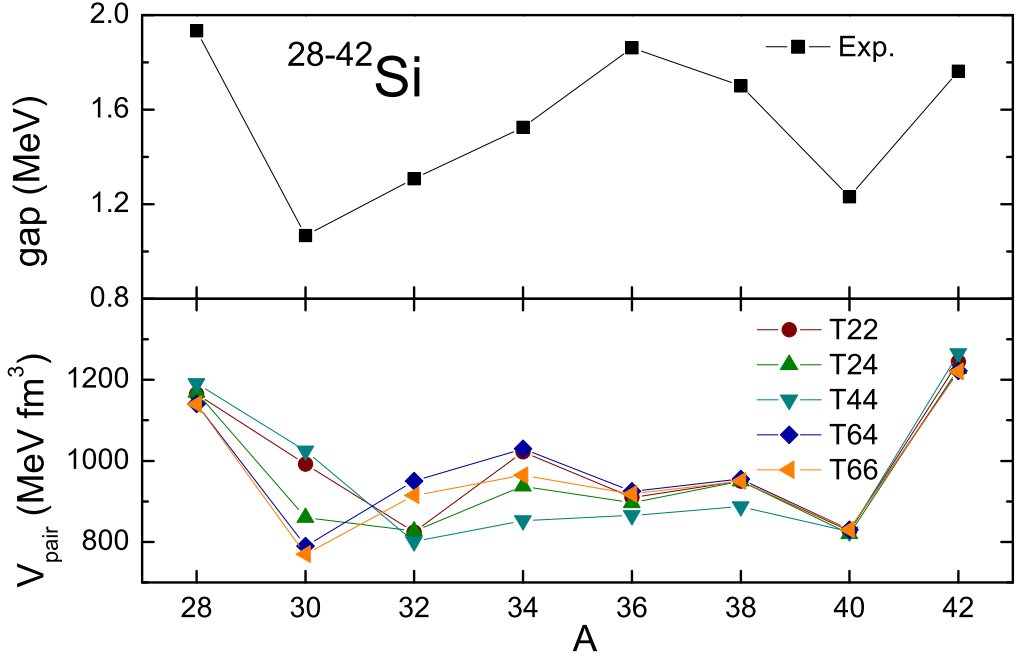}
\caption{(Color online) Upper panel: empirical neutron pairing gaps
of Si nuclei $^{28-42}$Si; Lower panel: adopted pairing strengths
$V_{\text{pair}}$ corresponding to the Skyrme effective force T22,
T24, T44, T64, and T66.}\label{fig2}
\end{figure}

%Table_______________________________________________________
%\begin{widetext}
\begin{table}[!h]
\caption{Coupling strengths (in MeV) of various parameter sets used
in the work.} \label{tensor}
\vspace{5pt} \centering
\begin{tabular}{cccccc}
\hline
~~~&~~~~~T22~~~~~&~~~~~T24~~~~~&~~~~~T44~~~~~&~~~~~T64~~~~~&~~~~~T66~~~~~
\\ \hline
~~~$\alpha$   & 0 & 120  & 120 & 120 & 240
\\ \hline
~~~$\beta$    & 0 & 0    & 120 & 240 & 240
\\ \hline
~~~$\alpha_T$ & -90.6 & 24.7 &8.97&-0.246 & 113
\\ \hline
~~~$\beta_T$  & 73.9 & 19.4 &113 & 218 & 204
\\ \hline
\end{tabular}
\end{table}%.......................................................
%\end{widetext}
The pairing interaction is written as
\begin{eqnarray}
V_{\text{pair}}(\br_1,\br_2) = V_{\text{pair}}
\left(1-\frac{\rho(\br)}{\rho_0}\right) \delta(\br_1-\br_2).
\label{pair}
\end{eqnarray}
where $\rho(\br)$ is the Hartree-Fock (HF) matter density at ${\br}
= ({\br}_1+{\br}_2)/2$ and $\rho_0 = 0.16\;\rm fm^{-3}$. The pairing
strengths $V_{\text{pair}}$ for each Skyrme force are determined to
reproduce the empirical neutron pairing gaps in the corresponding
nuclei. The gap data and the chosen pairing strengths adopted for
$^{28-42}$Si are presented in the upper and lower panel of
Fig.~\ref{fig2}, respectively. As a result the pairing strengths for
all the chosen effective forces tend to have an average value around
900 $\sim$ 1000 MeV fm$^3$ with variations up to 200 MeV fm$^3$.

%-------------------------------------------------------------------------------
\section{Results}
\begin{figure}
\centering
\includegraphics[width=13.5cm]{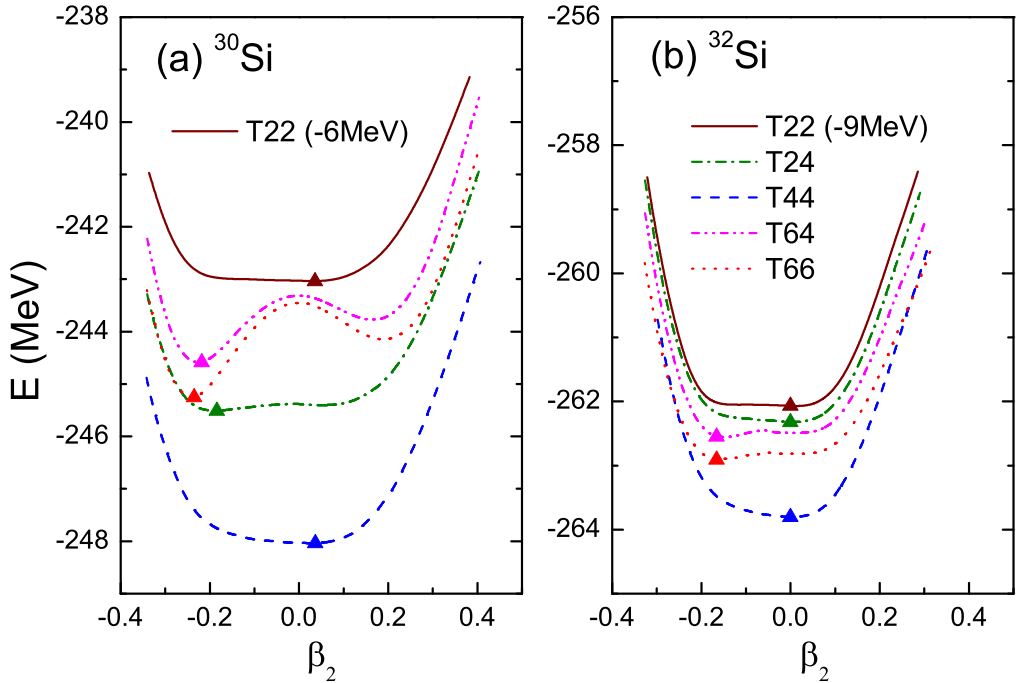}
\caption{(Color online) Energy surfaces of $^{30}$Si (left panel)
and $^{32}$Si (right panel) as a function of the quadruple
deformation parameter $\beta_2$ using T22, T24, T44, T64, and T66.
The results with T22 are shifted by $-$6 MeV and $-$9 MeV in
$^{30}$Si and $^{32}$Si, respectively. The energy minima are
indicated with triangles.}\label{fig3}
\end{figure}
We display in Fig.~\ref{fig3} the energy surfaces of $^{30}$Si (left
panel) and $^{32}$Si (right panel) as a function of the quadruple
deformation parameter $\beta_2$ using T22, T24, T44, T64, and T66.
The energy minima are indicated with triangles. To facilitate the
comparison, we shift the energy surfaces with T22 by a constant
amount as indicated in the figures. $^{30}$Si is suggested to be
deformed as mentioned in the introduction, but T22 and T44 with
relatively large pairing strengths ($\sim 1000$ MeV fm$^3$) fail to
give deformed energy minima. On the contrary, deformed ground states
can be achieved using the T24, T64 and T66 parametrizations with
resulting small pairing strengths ($\sim 800$ MeV fm$^3$). The
predicted oblate shape of this nuclei is consistent with the recent
RMF result~\cite{Li11}. One should note that the same gap data have
been used to deduce the pairing parameters for the different Skyrme
interactions, and the variation of pairing parameters is only aimed
to reproduce the experimental data. The interesting performance of a
weak nucleon pairing may stem from a well-known fact that the
pairing interaction forms the $J = 0^+$ pairs of identical particles
which have spherically symmetric wave functions, and as a result,
nuclei tend to be more spherical in the cases of T22 and T44 with
stronger pairing couplings. This suggests that the resulting shape
of a nucleus is sensitive to the nuclear pairing in each
nucleus~\cite{Li13}, and the experimentally-determined deformed
shape of $^{30}$Si favors relatively weak paring correlations
present in the Skyrme effective force T24, T64 and T66. We also
notice that large tensor terms present in T64 and T66 tend to make
the energy surface deep, namely the tensor force affects
dramatically to create a well-deformed ground state for $^{30}$Si.
The importance of tensor force will be seen also in $^{32}$Si as
shown in the right panel of the same figure. Its shape is predicted
to be spherical with T22, T24 and T44, but oblate with T64 and T66
with large tensor terms.

\begin{figure}
\centering
\includegraphics[width=13.5cm]{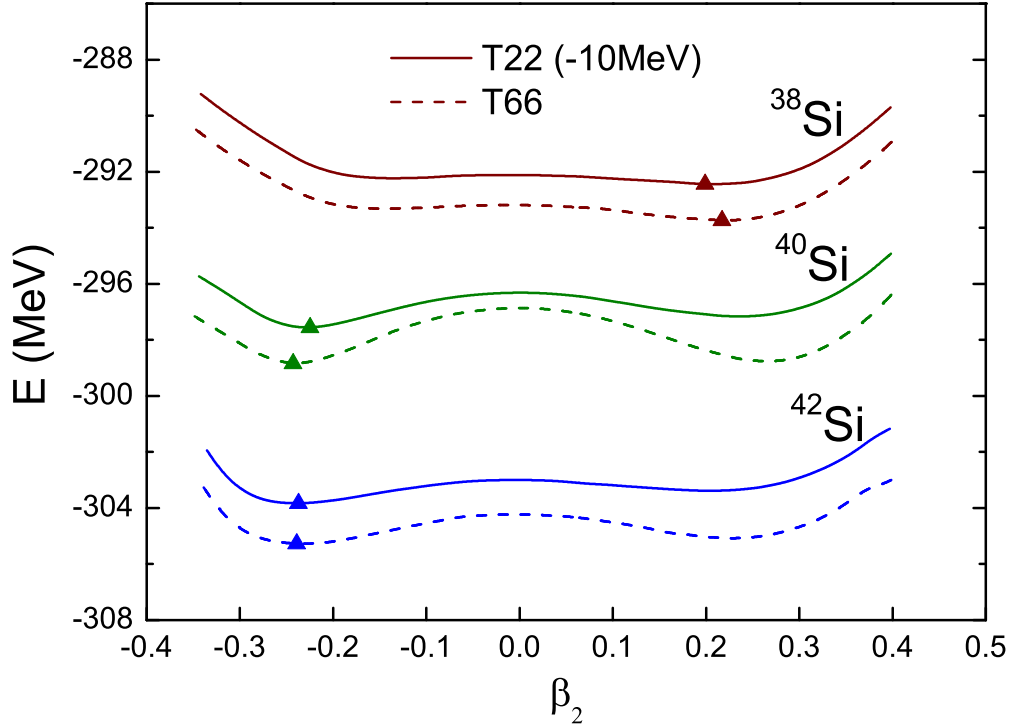}
\caption{(Color online) Energy surfaces of $^{38-42}$Si as a
function of the quadruple deformation parameter $\beta_2$ using T22
and T66. The results with T22 are shifted by $-$10 MeV. The energy
minima are indicated with triangles.}\label{fig4}
\end{figure}
Before proceeding to study in details the tensor role, we prepare in
Fig.~\ref{fig4} the energy surfaces of $^{38-42}$Si with two forces
T22 and T66, as two representatives to show the common features
among the five parameter sets in the present model. It is clearly
shown that the oblate shape of $^{42}$Si is quite robust for
different tensor couplings, which justifies the broken of $N = 28$
shell closure seen in a simple single-particle (s. p.) shell-model
picture. Also, large deformations are expected for $^{38,40}$Si from
their measured large $B(E2)$ values between the $2_1^+$ and the
$0_1^+$ states~\cite{Bas07,Ibb98}. The opposite sign of the
deformation parameter $\beta_2$ for $^{40}$Si was predicted by the
RMF model and the FRDM model~\cite{Wer96}. Experimentally the sign
of deformation will be determined by the measurement of its
quadrupole moment, and we do not draw any conclusion about this in
the present work. One more interesting feature is the deformation
change from $^{38}$Si to $^{42}$Si with the increase of neutron
number: the energy minima move from the prolate side in the lighter
nucleus $^{38}$Si to the oblate side in the heavier nucleus
$^{42}$Si. We should notice that this feature of shape evolution
will always appear in the present model, despite the employed Skyrme
effective force.

\begin{figure}
\centering
\includegraphics[width=13.5cm]{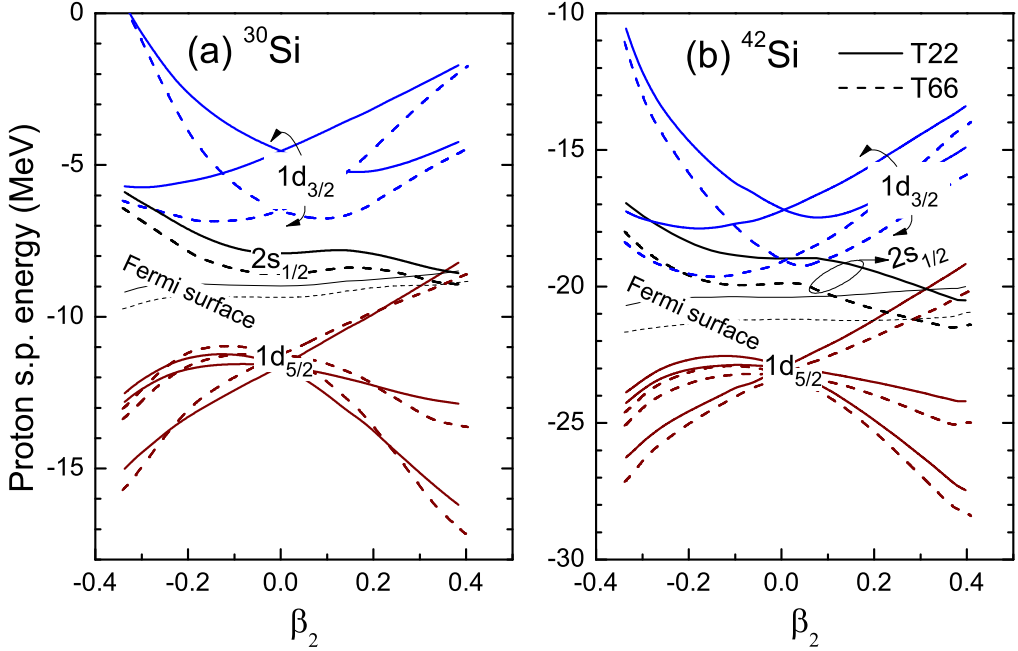}
\caption{(Color online) The s. p. energy levels of protons for
$^{30}$Si (left panel) and $^{42}$Si (right panel) are shown as a
function of the quadruple deformation parameter $\beta_2$ using T22
(solid lines) and T66 (dashed lines). The Fermi surface is also
shown between $2s_{1/2}$ and $1d_{5/2}$ orbits in the spherical
limit.}\label{fig5}
\end{figure}
Since the tensor force operates strongly between neutrons and
protons, the isotope dependence of tensor correlations could be
manifested in the difference of the proton s. p. orbits between
$^{30}$Si and $^{42}$Si. We present in Fig.~\ref{fig5} the single
proton levels in both $^{30}$Si (left panel) and $^{42}$Si (right
panel) as a function of the quadruple deformation parameter
$\beta_2$ using T22 (solid lines) and T66 (dashed lines). We choose
T22 and T66 to compare two extreme cases, i.e., T66 with the largest
tensor terms and T22 without the tensor terms. The Fermi surface is
also shown between $2s_{1/2}$ and $1d_{5/2}$ orbits in the spherical
limit. Since $\alpha$ and $\beta$ are positive in Eq. (\ref{Wq}) for
T66 case, the protons and the neutrons in $1d_{5/2}$ orbit decrease
the proton spin-orbit splitting in $^{30}$Si as shown in the left
panel of Fig. \ref{fig5}. Two main features can be seen as the
tensor effect to compare the results between T22 and T66 for
$^{30}$Si: a narrower $1d_{3/2}$-$1d_{5/2}$ gap and a steeper
$1d_{5/2}$ orbits downward in the oblate side. Then, in the case of
T66, more mixing between $1d_{3/2}$ and $1d_{5/2}$ results in an
oblate shape for this nuclei in comparison with the case of T22. The
shell gap by T66 interaction is narrower also in $^{42}$Si due to
the protons in the $1d_{5/2}$ orbit and the neutrons in $1f_{7/2}$
orbit as seen in the right panel of Fig. \ref{fig5}. However, both
T22 and T66 give oblate deformations, since the tensor effect in
this nuclei is not as evident as the case of $^{30}$Si regarding the
$1d_{5/2}$ orbits.

\begin{figure}
\centering
\includegraphics[width=13.5cm]{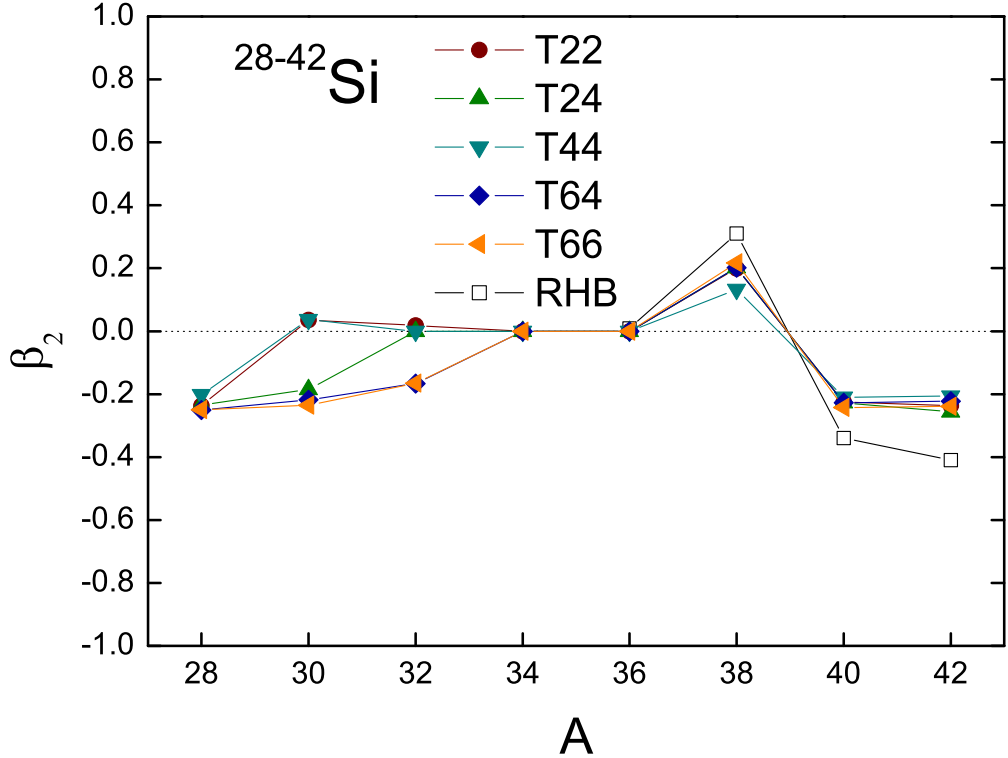}
\caption{(Color online) Deformation parameters $\beta_2$ as a
function of the mass number in comparison with the RHB
results~\cite{Lal99}. The calculated values corresponding to the
Skyrme effective force T22, T24, T44, T64, and T66.} \label{fig6}
\end{figure}

Finally we collect in Fig.~\ref{fig6} the evolution of the quadruple
deformations $\beta_2$ with mass number $A$ along the silicon
isotopic chain for various effective interactions, in comparison
with the Relativistic-Hartree-Bogoliubov (RHB) results~\cite{Lal99}.
In general, the deformation evolution with increasing neutron number
demonstrates clearly the typical Jahn-Teller effect~\cite{Naz93}:
centered with a spherical shape for $^{34}$Si nucleus at a closed
shell, the isotope chain evolves from an oblate shape to a prolate
shape. All parametrizations in Fig.~\ref{fig6} predict an oblate
shape for $^{28,40,42}$Si, a spherical shape for $^{34,36}$Si, and a
prolate shape for $^{38}$Si. The spherical shape of $^{34}$Si is
expected because of the neutron closed shell $N = 20$. The same
spherical shape might be also expected in $^{36}$Si because two
neutrons outside of the $N = 20$ closed shell is not enough to make
a deformation. However, the tensor force has a large effect in
nuclei $^{30,32}$Si deepening the energy surfaces at oblate energy
minima in these nuclei. This mechanism can be understood from the
shell gap narrowing due to the tensor force showing in the left
panel of Fig.~\ref{fig5}.

\section{Conclusion}
In summary, we have applied the DSHF model to study the tensor
effect on the evolution of the quadruple deformation in Si isotopes,
using the experimentally-determined pairing strengths. We found a
manifestation of tensor correlations in several isotopic nuclei,
such as $^{30,32}$Si, in competition with the pairing correlations.
Generally, the tensor force helps to achieve a well-deformed local
minimum, by reducing the shell gaps and enhancing s. p. level
densities near the Fermi level. In our calculations, the effect of
tensor force emerges evidently when we compare the calculated shapes
of $^{32}$Si with increasing tensor terms in the employed HF energy
density. Large tensor correlations result in a larger deformation
for this nuclei, in comparison with a spherical result with small
tensor forces. The tensor-force-driven deformation in this nuclei is
something we should pay much attention to, since it relates to a
further improvement of many theoretical models or parametrizations,
such as shell model and the SHF model, toward a better description
for the shell structures of nuclei in general. On the other hand,
the pairing interaction among nucleons is introduced in this work
using the BCS treatment. We take the strength parameters fully
consistent with the empirical gap data. For the chosen various
Skyrme parametrizations, we find that the resulting pairing
strengths are not so much varied around the average value around
$V_{\text{pair}}$ = 900 $\sim$ 1000 MeV fm$^3$. However, the
resulting nuclear shapes are sensitive to the adapted pairing
strengths. That is, to take $^{30}$Si as an example, T22 and T44
with the larger values around 1000 MeV fm$^3$, give no deformation,
while T24, T64, and T66 with the smaller values around 800 MeV
fm$^3$ give a large oblate deformation. We notice that the latter
case is preferred by the recent experiments. This means the Skyrme
interactions T24, T64, T66 give a better description of $^{30}$Si.
Finally, all the listed effective forces in this work show a
prolate-to-oblate transition with increasing number of neutrons from
$^{38}$Si to $^{42}$Si as a typical Jahn-Teller effect. To further
disentangle the prolate or oblate nature of $^{40}$Si, the Coulomb
excitation could be the best experimental tool to find out the sign
and the magnitude of the quadrupole deformation. The measurements of
quadruple moments of its neighboring nuclei (such as $^{39}$Si or
$^{41}$Si) can also give important information on the evolution of
the deformation near $^{40}$Si. We should also notice that these
observables are affected by the polarization effect which is another
interesting issue to be studied.

\section*{Acknowledgment}
We would like to thank E. Hiyama for valuable discussions. This work
is supported by the National Natural Science Foundation of China
(Grant Nos. 10905048, 10975116 and 11275160).

%-------------------------------------------------------------------------------
\bibliographystyle{plainnat}

\end{document}